# Formation of nanostructures in ferroelectrics and antiferroelectrics in the process of phase transformation

V.M. Ishchuk

Institute for Single Crystals NAS of Ukraine, 61001 Kharkov, Ukraine

N.A. Spiridonov

Science and Engineering Center "Reaktivelektron", NAS of Ukraine, 83096 Donetsk, Ukraine

V.L. Sobolev

Department of Physics, South Dakota School of Mines and Technology, Rapid City, SD 57701

This paper contains results of investigations of inhomogeneous states caused by the coexistence of ferroelectric and antiferroelectric phases in lead-zirconate-titanate based solid solutions. The domains of ferroelectric and antiferroelectric phases with sizes of the order of 20 to 30 *nm* coexist in the bulk of the samples due to a small difference in the free energies of the phases. The coherent character of interphase boundaries leads to the concentration of elastic stresses along these boundaries. Elastic stresses cause the local decomposition of the solid solution due to the circumstance that equivalent positions of the crystal lattice are occupied by ions with different sizes. Larger ions are driven out into domains with larger crystal cell parameters and smaller ions are pushed into domains with smaller parameters of the crystal cell. The sizes of segregates formed in this way are of the order 8 to 15 *nm*.

77.84.-s, 77.80.Bh, 77.80.Dj

# 1. Introduction

Solid solutions of complex oxides with perovskite crystal structure have been the focus intense fundamental studies for last forty years due to their importance for a wide variety of applications[1,2,3,4,5]. The technological interest arises from the fact that these materials may be prepared with very high values of dielectric, piezoelectric, electrostrictive, and electrooptic constants in wide temperature ranges, finding applications like actuators, nonvolatile and dynamic random access memories, capacitors, microelectromechanic devices, electrooptic modulators, optical switches, etc.

Lead lanthanum zirconate titanate (PLZT) solid solutions $Pb_{1-3x/2}La_x(Zr_{1-y}Ti_y)O_3$ possess wide intervals of thermodynamic parameters in which the domains of the ferroelectric (FE) and antiferroelectric (AFE) phases coexist due to the small difference in free energies of FE and AFE states. The inhomogeneous state of the coexisting phases was studied by several groups of researchers[5,6,7,8,9,10,11,12,13,14,15,16,17,18,19]. It was found that this inhomogeneous state is stable with respect to homogeneous states of each phase. The last circumstance leads to a number of remarkable effects[5-8,10,15,16]. PLZT solid solutions with La content $x \geq 6\%$ are of significant interest in connection with investigations of the nature of the so-called relaxor state[20,21,22,23,24,25].

In this work the results of studies of inhomogeneous states caused by coexistence of ferroelectric and antiferroelectric phases in lead-zirconate-titanate (PZT) based solid solutions and formation of nanostructures caused by local decomposition of these solid solutions are presented.

The following systems of solid solutions were chosen for our studies: PLZT with La-content $x = 0.06$, and $Pb_{1-x}(La_{1/2}L_{1/2})_x(Zr_{1-y}Ti_y)O_3$ (PLLZT) with $x = 0.10, 0.15$. In what follows we are going to use the following symbols for solid solution compositions X/100-Y/Y, where X and Y show the percentage content of substituting elements. The "composition – temperature" phase diagrams (Y-T phase diagrams) for both systems of solid solutions are similar. The Y-T phase diagrams for PLLZT are given in Fig. 1. The regions of FE and AFE states are separated by a wide crosshatched region (border region) where domains of FE and AFE phases coexist. Such coexistence of phases has been identified by X-ray methods[5-8] in both PLZT and PLLZT and coexisting domains were directly observed by means of transmission electron microscopy in the PLZT solid solutions[5,26]. The FE phase of both systems of solid solutions is characterized by the rhombohedral type of lattice distortions and the AFE phase possesses tetragonal lattice distortions. The border region in the phase diagrams in the Fig. 1 has no sharp boundaries. The region of induced FE states appears on the Y-T phase diagram after the sample was subjected to the action of the electric field. The evolution of the Y-T phase diagrams after the action of the DC electric field on the three series of PLZT solid solutions is presented in Fig. 2. The diagrams acquire a more complex structure[6-8] but these effects will not be considered in this paper. The intermediate region of coexisting states occupies the interval of Ti-compositions from 20% to 29% in PLZT solid solutions.

## 2. Experimental Methods

The initial batch with the necessary PLLZT composition was obtained by standard ceramic technology with two stage synthesis at 850 °C and then at 1000°C. The initial batch with the PLZT composition was obtained by mutual co-precipitation from the mixture of aqueous solutions of lead and lanthanum nitrates and chlorides of titanium and zirconium with a subsequent two stage synthesis at 550 °C and 850 °C. Polycrystalline samples were obtained by sintering at the temperatures 1300 °C – 1350 °C in a controlled PbO atmosphere.

Investigations of the sample crystal structures were carried out by means of a SIEMENS D-500 powder diffractometer using a Ge monochromator ($CuK_{\alpha 1}$ radiation with a wave length of 1.54056 Å) and a BRAUN gas position sensitive detector. The accuracy of the 2θ angle measurements was 0.01°.

Investigations of mesoscopic structure of the solid solutions were carried out by Debye's method (on the sample's metallographic section) with subsequent photometry of X-ray patterns. The $CoK_\alpha$ radiation filtered by a layer of vanadium oxide was used. The angular speed of a flat sample was 1 *rpm*. The duration of the survey was 20 *min* (10 *min* for each position of the sample plane, symmetric with respect to the incident X-ray beam).

Measurements of dielectric and piezoelectric characteristics were carried out by standard methods.

# 3. Experimental results

## 3.1. X-ray studies in paraelectric phase

Investigated solid solutions with compositions belonging to the region of coexisting FE and AFE phases (the dashed region in the Y-T phase diagram in the Fig. 1 and Fig.2) are characterized by a highly diffused phase transition from paraelectric (PE) to dipole ordered phase. The degree of diffuseness[27] (characterized by the parameter $\delta$) decreases considerably when the content of titanium is greater or smaller then the concentrations corresponding to the above-mentioned region. The dependence of the $\delta$-parameter on the titanium composition $\delta(Y)$ is shown in Fig.3 (Curve number 1).

We investigated temperature dependencies of crystal lattice parameters and shapes of the X-ray lines. The (222) and (400) X-ray diffraction lines were chosen for the investigations. These lines are the most characteristic ones for the substances in question. The (222) X-ray diffraction line is a singlet in the case of tetragonal distortions of the crystal lattice and becomes a doublet when the lattice distortions are of the rhombohedral type. The (400) X-ray line is a singlet in the case of rhombohedral distortions and is a doublet for the case of the tetragonal distortions of the perovskite crystal lattice. The smearing of the phase transition from the paraelectric phase into the dipole ordered phase is connected with the existence of domains of ordered phases above the point of the paraelectric phase transition. Due to this circumstance the analysis of the line shape allows one to draw conclusions regarding the types of domains of the ordered phases that exist in the paraelectric matrix of the substance at high temperatures.

Experimental results on the temperature dependencies of X-ray lines are summarized in the Fig.4. Fig.4 presents the temperature dependencies of the parameter $\gamma$, which characterises the asymmetry of the X-ray lines. This parameter is a ratio $\gamma = S_l/S_r$ (see Fig. 4a) where $S_l$ and $S_r$ are the areas below the contour of the X-ray diffraction line located from the left and from the right of the vertical line drawn from the vertex of the X-ray line. The total area of the X-ray diffraction line is equal to $S = S_l + S_r$. Dependencies of the asymmetry parameter $\gamma$ and crystal lattice parameters for 10/100-Y/Y PLLZT solid solutions are given in Fig. 4b. The solid solution 10/90/10 and 10/70/30 correspond to materials whose compositions are located outside the dashed region in the Y-T phase diagram (see Fig.1) whereas the 10/80/20 solid solution is inside this region. As one can see both (222) and (400) lines for the samples with compositions from the dashed region of the Y-T phase diagram are asymmetric in a wide interval of temperatures in the PE phase. This asymmetry is the consequence of the existence of the two-phase (FE+AFE) domains in the paraelectric matrix of the substance[28,29] at temperatures $T > T_C$ (see also section 4.1).

The two-phase domains also present at the temperatures below $T_C$ in the solid solutions with compositions from the region of coexistence of the FE and AFE phases. The existence of these domains is important for understanding the mesoscopic structure of solid solutions at the temperatures below $T_C$. Analysis of the complex character of profiles of the (200) and (222) X-ray diffraction lines (these are the most characteristic lines for examination of the perovskite crystal structure) and their intensities for the solid solutions with different compositions in the vicinity of the region of coexistence of FE and AFE phases (see Fig.2) as well as from this region in the Y-T phase diagram[30] allowed to obtain the dependence of the fraction of the FE phase in the sample volume as a function of Ti content. This dependence is given in the Fig.5. The position of the peak of each component of the complex X-ray diffraction line depends on the Ti content in the solid solution. The dependencies of these positions on Ti content for the (200) and (222) X-ray lines along with the dependencies of the crystal cell parameters are presented in the Fig.6 to illustrate the change of the solid solution crystal structure with the variation of composition.

**3.2. Piezoelectric studies in the paraelectric phase**

The presence of two-phase domains, which contain the FE component, allows one to use the piezoelectric resonance method for investigations of phase transition. The said method cannot be applied at temperatures $T > T_C$ when the FE component is absent (provided that the external electric field is also absent).

In our studies the experiment was carried out for two different regimes of thermoelectric treatment of the samples. In the first case the samples were annealed at 500 °C for 1 hour and after that they were cooled at the rate of 4 degree/min. An electric field (800 *V/mm*) was applied to the samples at 300 °C (which is above the Curie point) and the cooling continued in the presence of the field until the temperature reached $T_C$ + 20 °C, at which the electric field was switched of and the samples were cooled down to room temperature. Measurements of piezoresonance were carried out at room temperature. In the second case the experimental conditions were slightly different. After annealing the electric field was applied to the samples at $T_1 = T_C + 20$ °C (the condition that $T_1 > T_C$ was observed all the time). The samples were kept at this temperature and the applied electric field for 0.5 hour, after that the field was switched off and the samples cooled down to room temperature and the measurements were carried out.

Dependencies of the piezoelectric coefficient $d_{33}$ on the Ti-content of the PLLZT solid solutions of the system 10/100-Y/Y for the two above-described regimes of the thermal treatment are shown in Fig. 3 (curves 2 and 3 refer to the first and second regime respectively). Both dependencies have well pronounced maxima located near Y = 20 (which corresponds to 20% Ti). Solid solutions do not manifest piezoelectric properties when the Ti-content moves out from this value.

Analogous results are obtained for the PLZT solid solutions.

### 3.3. Long-term relaxation

The mesoscopic structures of the solid solutions under investigation is related to coexistence of domains of FE and AFE phases at temperatures below Curie point [31] and is caused by local segregations in the vicinity of interphase boundaries. The main task of our investigation is to trace the kinetics of formation of such structures. One can confirm this fact experimentally by subjecting the solid solution with composition from the region of the Y-T phase diagram that corresponds to the inhomogeneous structure of coexisting FE and AFE phases to quenching from high temperatures when this solution is in the PE state. The time required for the sample to establish an equilibrium state with a definite value of order parameters is typically from $10^{-6}$ to $10^{-4}$ seconds whereas the formation of segregates is caused by diffusion processes that are long-term processes at room temperatures.

Samples of solid solutions were annealed at 600 °C for 22 hours. After that the samples were quenched to room temperature. The samples were aged during some interval of time $\tau$ and X-ray studies were carried out. Diffraction patterns that were obtained at 600 °C (after the annealing) are characterized by the presence of singlet Debye's lines only. The structure of the diffraction patterns becomes more complicated in the process of the sample's aging. Broadened diffuse lines (haloes) that have significantly less intensity appear in addition to Debye's lines. These haloes appear in the interval of angles $\theta = 25° - 27°$ (halo 1) and $\theta = 29° - 32°$ (halo 2). The intensity, location, and shape of the halo change with time. The shape and location of Debye's lines, which characterize the crystal structure of the solid solution under investigation also change.

The dependence of the elementary cell volume on the aging time for the PLZT solid solution with composition 6/73/27 is given in Fig. 7 as an example. The shape, intensity and position of the halo as functions of aging time are presented in Fig. 8. Changes of the shape and position of the (111) and (200) Debye lines in the process of sample aging is given in the Fig. 9.

## 4. Discussion of experimental results

### 4.1. Two-phase nucleation

Analysis of the profiles of the lines in diffraction pattern shows that the solid solutions in question have complex composite structures at temperatures above the Curie point. In the first approximation such a structure may be considered as the case when two-phase (FE+AFE) domains, which can be considered as joint domains of FE and AFE phases, coexist in the paraelectric matrix of the sample. Such structure can be understood from the following reasoning. The volume of the elementary crystal cell in ferroelectrics increases as compared to the volume of the elementary crystal cell in the PE phase during the phase transition into the ordered state. The volume of the elementary crystal cell decreases when the phase transition into the AFE state takes place. Thus, if domains that existed in the low temperature phase (the phase at $T < T_C$) continue to exist in the bulk of the sample above the point of phase transition it might lead to an increase of elastic energy (as a consequence of the difference in configuration volumes of low-temperature and high-temperature phases). In the case when the two-phase (FE+AFE) domains appear its configuration volume does not differ from the configuration volume of the PE phase and as a result the elastic energy of the sample as a whole will not increase. The ratio of volumes of the FE and AFE phases inside of such two-phase domains depends on the relative stability of the FE and AFE states for each particular solid solution. The relative stability in turn depends on the position of this solid solution in the Y-T phase diagram (the solid solution composition) with respect to the boundary separating the regions of the FE and AFE states). This ratio also depends on the difference of configuration volumes of the PE state and configuration volumes of each of the ordered states. Additional stabilization of such two-phase states in the PE matrix of the sample is provided by the interaction between FE and AFE phases inside the domain[10,16]. This interphase interaction is caused by long-range fields generated by the order parameters of the phases. In particular, such interaction provides the stability of two-phase states at the low temperatures (crosshatched region in the Y-T diagram in the Fig. 1 and Fig. 2).

TEM images of the coexisting domain of the FE and AFE phases were obtained using single crystallite samples that were produced by cleavage of coarse-grained PLZT ceramics with the 7/65/35 composition (the grains with the sizes of about 10 $\mu m$ were selected for these experiments). The samples obtained as result of cleavage had the shape of platelets plates with the thickness of 0.1 – 0.2 $\mu m$ and (110) crystal plane was the plane of the platelets. TEM images of two orientations of the (110) cleavage plane of a single crystallite sample with respect to the electron beam in Fig. 10 show the AFE matrix containing the domains of the FE phase.. The left photograph demonstrates the bright-field image of the (110) cleavage plane. A clear two-phase structure is observed inside the crystallite in this case. The second-phase inclusions shaped as oval have dimensions of the order of $(2 - 4) \cdot 10^{-6}$ $cm$ and a density of the order of $10^{11}$ $cm^{-2}$. The right photograph in the Fig. 10 presents the image of the (120) plane of the same part of the sample. In this case a "fibrous" two-phase structure manifests itself. Analysis of the images obtained at different sample orientations with respect to the electron beam shows that the second-phase inclusions have cylindrical shape and penetrates through the whole depth of a thin crystallite in the case of (110) cleavage. The "fibrous" structure seen in Fig. 10 is a consequence of the electron beam propagation through a thin crystallite under an angle with respect to the axis of cylindrical inclusions.

FE and AFE phases have different configuration volumes; therefore, their coexistence may be accompanied by the formation of dislocations along interphase boundaries (IPB) or by the emergence of elastic stresses. No dislocations along the FE-AFE interphase boundaries were found. Thus, IPBs separating these phases have a coherent character. Photographs did not shown typical manifestations of matrix distortions caused by elastic stresses, too which implies that there is no concentration of deformation in the vicinity of IPBs. It was determined that the sizes of these domains are of the order of 20 – 30 *nm*.

### 4.2. Local decomposition of the solid solution

The information about the IPBs separating domains of AFE and FE phases presented in previous section allows us to conclude that the crossing through the region between two domains or in other words crossing through the IPB (the boundary in question can be named as seed or better to say "bare" IPB) is accompanied by a continuous conjugation of crystal planes. Such coherent character has to be accompanied by an increase of elastic energy of the crystal lattice along the boundaries.

The equivalent positions of the crystal lattice in the substances under consideration are occupied by ions with different sizes and/or different electric charges. In the bulk of each domain (away from the domain boundaries) the net force acting on each of these ions is equal zero. In the vicinity of IPB the balance of forces is disturbed, "Large" ions are driven out into domains with larger configuration volume and correspondingly with larger distances between atomic planes. "Small" ions are driven out into domains with smaller distances between atomic planes. Such processes are accompanied by a reduction in elastic energy along IPB on the one hand, and an increase of the energy caused by the deviation of the solid solution composition from the equilibrium composition. The process described above will be finished when the structure of the new IPB will correspond to the minimum of energy. This new "dressed" IPB is not a bare IPB any more. In what follows we will use the term IPB for the "dressed" interphase boundary. The A-positions of the perovskite crystal lattice in the solid solutions under consideration are occupied by ions with different ionic charges ($Pb^{2+}$, $La^{3+}$, $Li^+$). Because of this the local decomposition of the solid solution along the IPB can be accompanied by the local disturbance of electro-neutrality. Thus, the formation of heterophase structure is accompanied by a violation of the chemical homogeneity of the samples. The samples remain homogeneous at high temperatures when the coexistence of FE and AFE phases is absent.

### 4.3. Creation of textured nanostructures

One can control the microstructure of the solid solutions by the virtue of the fact that the inhomogeneous structure and segregates along the interphase boundaries appear in the process of the phase transition and the phase transition can be easily controlled. In particular one can create textured structure in the materials under consideration and such structures will remain stable after external influences on the samples are switched off. The results of our experiments on the excitation of piezoelectric resonance (section 3.2) are the example of such situation. If local decomposition of the solid solution does not take place in the samples, then these samples remain in a macroscopically depolarized state after the regimes of the thermal treatment described in section 3.2 and cooling down to room temperature. Piezoelectric modules of these macroscopically depolarized samples are practically zero (see the data for solid solutions with Ti-content far from 20% in the Fig. 2). The situation is quite different for solid solutions with approximately equal stability of the FE and AFE states. The two-phase domains that appear in the PE matrix of the samples in the process of cooling down from high temperatures have random distributions of spontaneous deformations and, correspondingly, random distributions of the axes of spontaneous polarization in the absence of an electric field. The distribution of the planes of interphase boundaries is also random. The process of diffuse segregation along the boundaries leads to the fixation of these random spatial distributions in the course of further cooling. The growth of the volume occupied by two-phase domains also takes place during the temperature decrease. Near $T_C$ the whole volume of the sample undergoes transition into an inhomogeneous state of coexisting FE and AFE phases with the directions of spontaneous polarization pinned by segregates. If one would try to polarize such a sample, the degree of polarization is going to be quite small and the level of piezoelectric properties will be minor.

In the case when the cooling of the sample from high temperatures is carried out in the presence of the electric field the distribution of the axes of spontaneous polarization (along with the distribution of directions of interphase boundaries) of nucleated two-phase domains will be given by the direction of the field. In this case texture is created in the samples after cooling. This texture is stabilized by the segregates on the interphase boundaries at room temperature (or temperatures close to room temperature). The samples are macroscopically polarized and it is easy to excite the piezoelectric resonance in these samples. The piezoelectric modules are nonzero even in the case when additional polarization (after the thermal treatment) of the samples has not been carried out.

If the additional polarization of the solid solutions by means of an electric field directed as in the process of prior thermoelectric treatment has been carried out then materials have the set of properties that cannot be achieved by means of any other treatments.

### 4.4. Long-term relaxation

The analysis of the time dependencies of the elementary cell volume, and the analysis of the time dependencies of the shape of diffuse halos, as well as the fact of the absence of halos in diffraction patterns obtained at 600 $^\circ C$ confirm one more time that the existence of the interphase boundaries is the cause of the long-term relaxation and the formation of segregates.

It is necessary to note that the process of establishment of an equilibrium state is the long-term process in solid solutions in which the state of coexisting FE and AFE phases is realized. As one can see from the X-ray data it continues not less then 120 hours. However, taking into account the sensitivity of this method one can assert that this process takes an even longer time.

The process is a multistage one. It is clearly seen from the results given in Fig. 4-6 that there are two relaxation times caused by different mechanisms. In addition to that one has to note the mechanism responsible for the establishment of equilibrium values of structural order parameters at the time intervals from $10^{-6}$ $s$ to $10^{-4}$ $s$ (such time intervals are beyond the abilities of our experimental methods). Without elucidation of particular mechanisms for establishing equilibrium state one can however argue that the long-term character of this process is connected with diffusion processes of the local decomposition of solid solutions along the interphase domain boundaries. The estimation of the size of the segregates (using the shape of the halo) gives the values of 8 *nm* to 15 *nm*.

Long-term relaxation processes are non-monotonous processes due to the condition of "strong deviation from equilibrium" at the initial stage right after quenching. In the case of "weak deviation from equilibrium" at the final stage the relaxation process is monotonous and is described by an exponential law. Peculiarities of the crystal structure of concrete solid solution also influence the relaxation process. In particular, the PLZT and PLLZT solid solutions differ by the presence of vacancies in the A-positions of the crystal lattice of the first system of solid solutions. The relaxation process follows similar patterns for both systems. However, there is a peculiarity of this process for the second system of solid solutions at aging times 20-30 hours, which we attribute to accumulation of elastic stress and their subsequent drop.

## 5. Conclusion

The domains of FE and AFE phases coexist in the sample volume due to a weak difference in free energies of FE and AFE states. These domains have sizes form 20 *nm* to 30 *nm*. Such domains coexist both at low ($T < T_C$) and at high ($T > T_C$) temperatures. In the last case they represent two-phase domains in the PE matrix of the substance.

Elastic stresses at coherent boundaries lead to the local decomposition of the solid solution along the interphase FE-AFE boundaries due to the circumstance that equivalent positions of the crystal lattice are occupied by ions of different size. As a result the "large" ions are ions are driven out into domains with larger parameters of the crystal lattice and "small" ions are driven into domains with smaller lattice parameters. The sizes of segregates formed in this process are of 8 *nm* to 15 *nm*.

The process of the establishment of equilibrium state of coexisting domains of FE and AFE phases is a long-term process and is determined by several different mechanisms.

One can control the process of the local decomposition of solid solutions by subjecting the substance to external influences. In particular, one can create textured structures that differ from ordinary piezoelectric materials by higher values of their parameters.

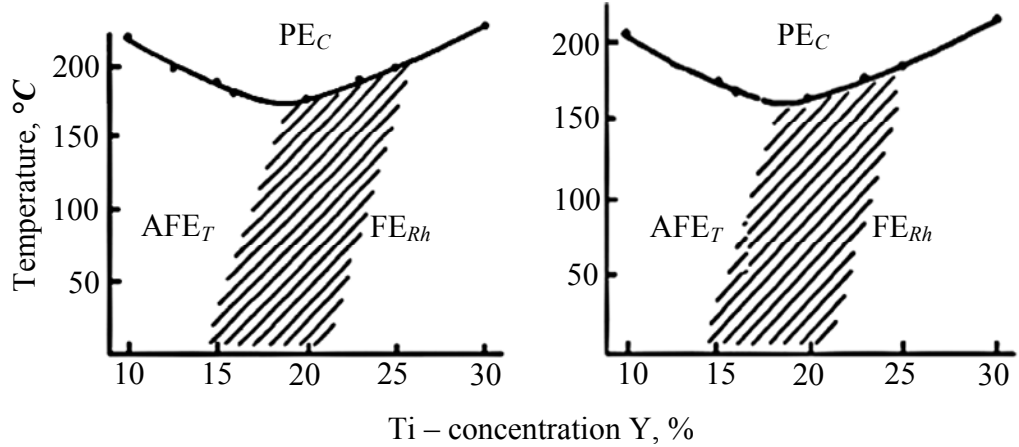

Fig.1. The Y-T phase diagrams for the series of PLLZT solid solutions 10/100-Y/Y (left) and 15/100-Y/Y (right). Here $PE_C$ stands for cubic paraelectric phase as well as $AFE_T$ and $FE_{Rh}$ stand for tetragonal antiferroelectric and rhombohedral ferroelectric phases, respectively.

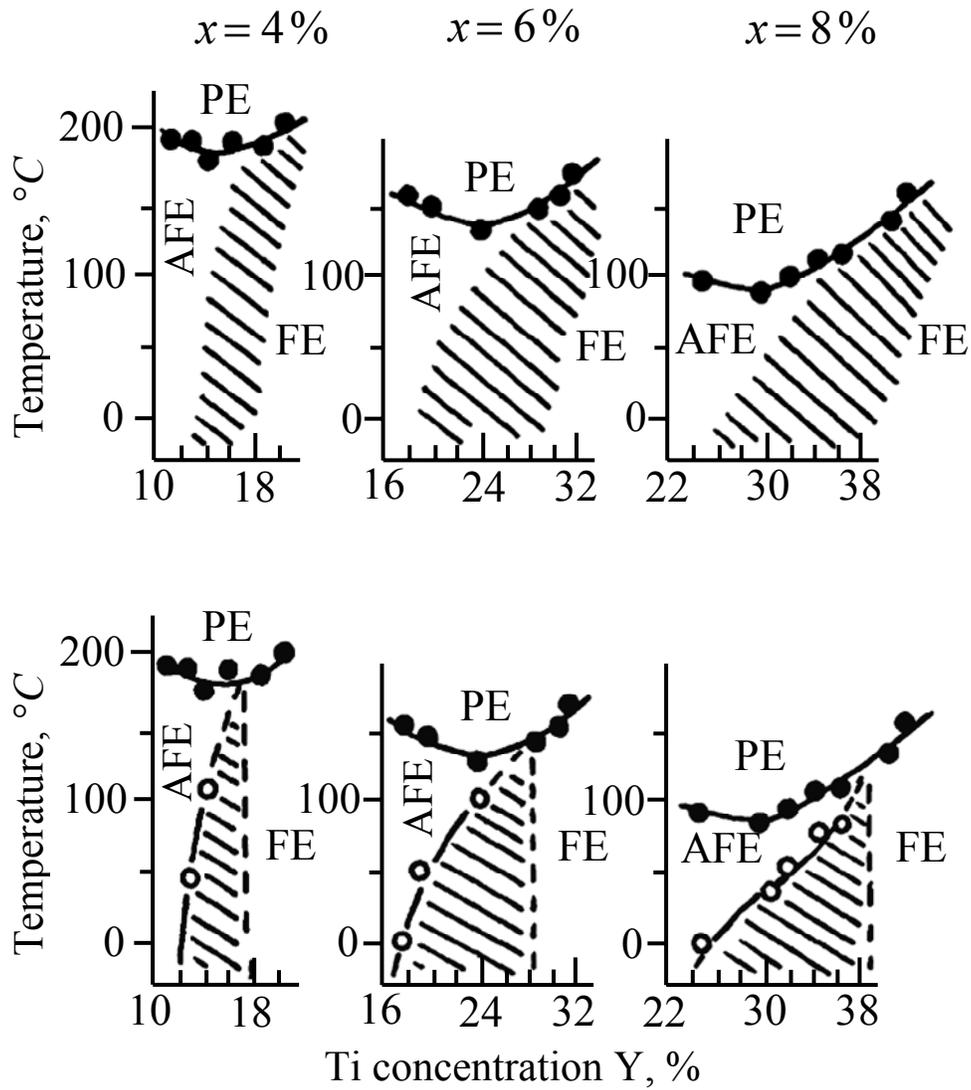

**Fig.2.** The Y-T phase diagrams of the PLZT solid solutions before (above) and after (below) the action of the DC electric field.

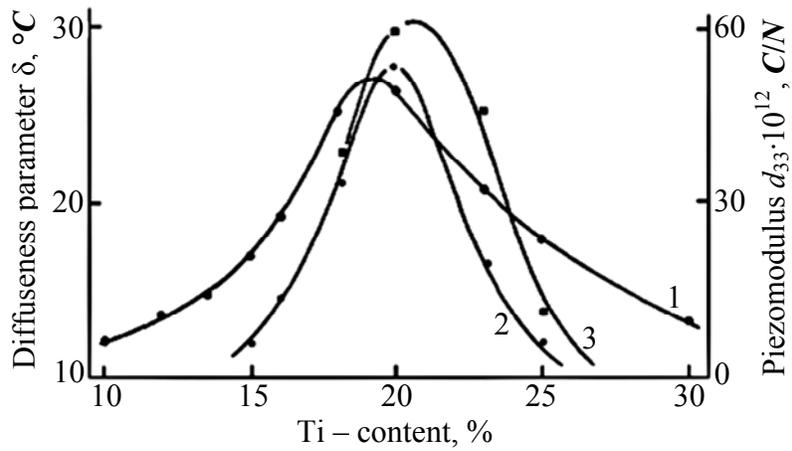

**Fig.3. Dependencies of diffuseness parameter δ (curve 1) and piezoelectric modules (curves 2 and 3) on composition for the 10/100-Y/Y series of PLLZT solid solutions measured after different regimes of thermal treatment in paraelectric phase (see part 3.2).**

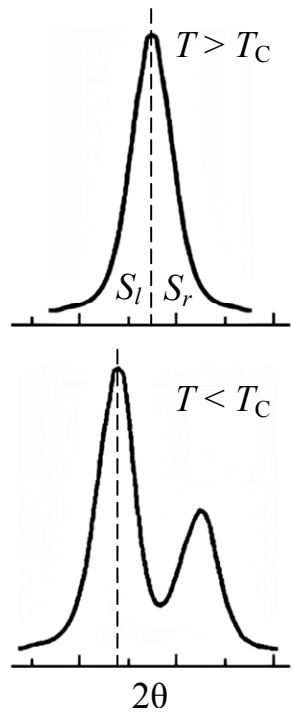

**Fig.4a. Illustration of parameter γ = S$_l$/S$_r$.**

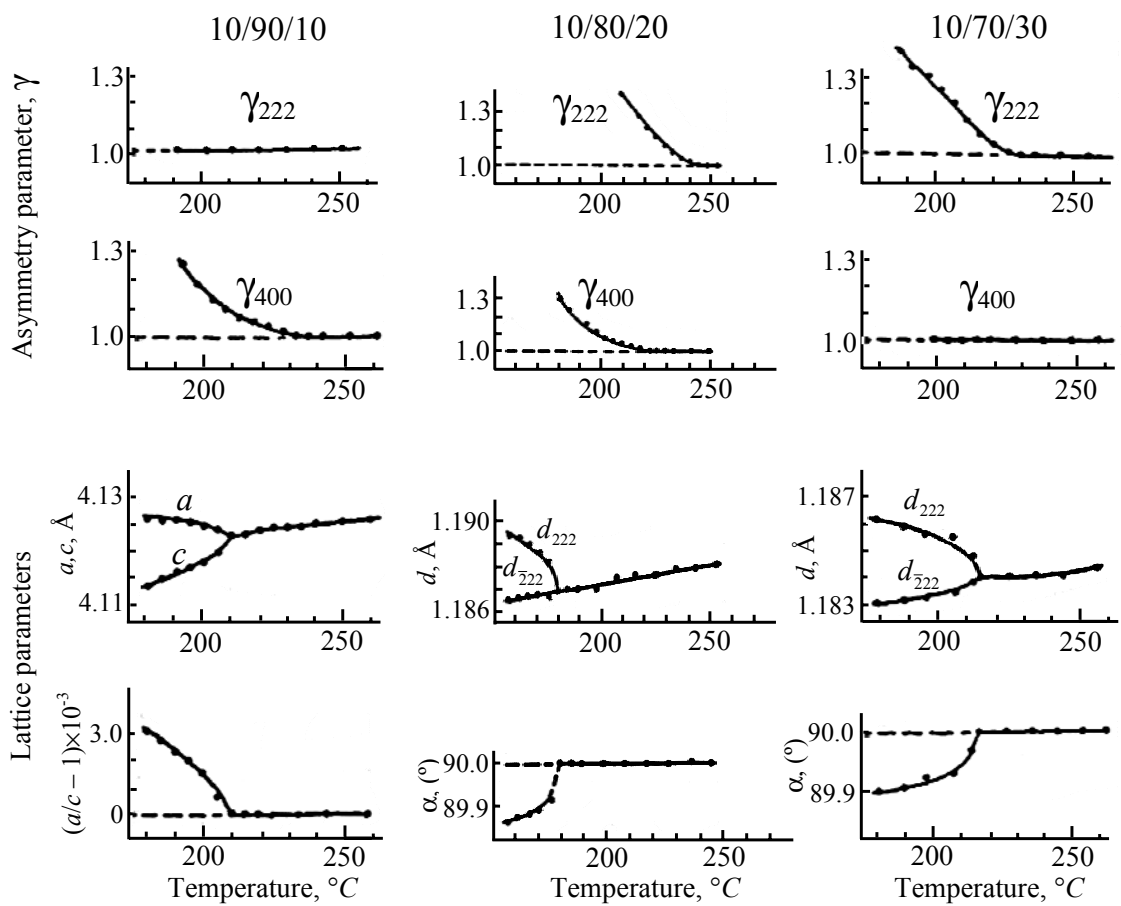

**Fig.4b.** Temperature dependencies of crystal lattice parameters for the 10/100−Y/Y series of PLLZT solid solutions. Here $\gamma_{hkl}$ is the asymmetry parameter of the (hkl) X-ray line; $d_{hkl}$ Interplane distances for the (hkl) family of crystal planes; *a* and *c* are the parameters of the perovskite crystal lattice in tetragonal phase, $(a/c - 1)$ is the degree of tetragonality; α is the rhombohedral angle.

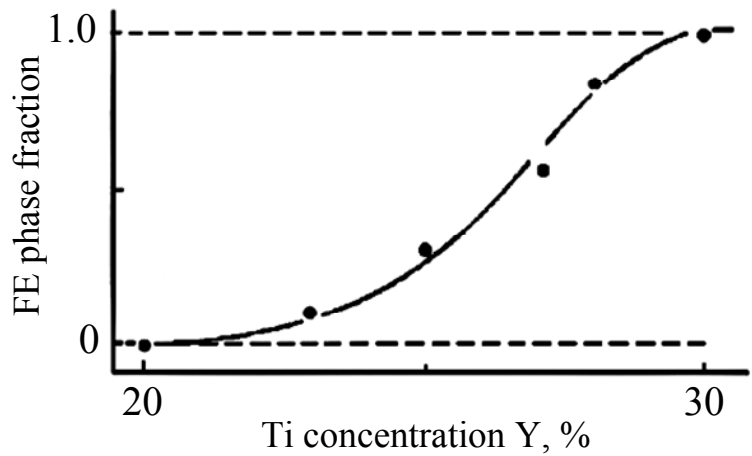

Fig.5. The dependence of the fraction of FE (rhombohedral) phase in the sample volume on the content of Ti in 6/100−Y/Y system of PLZT solid solutions.

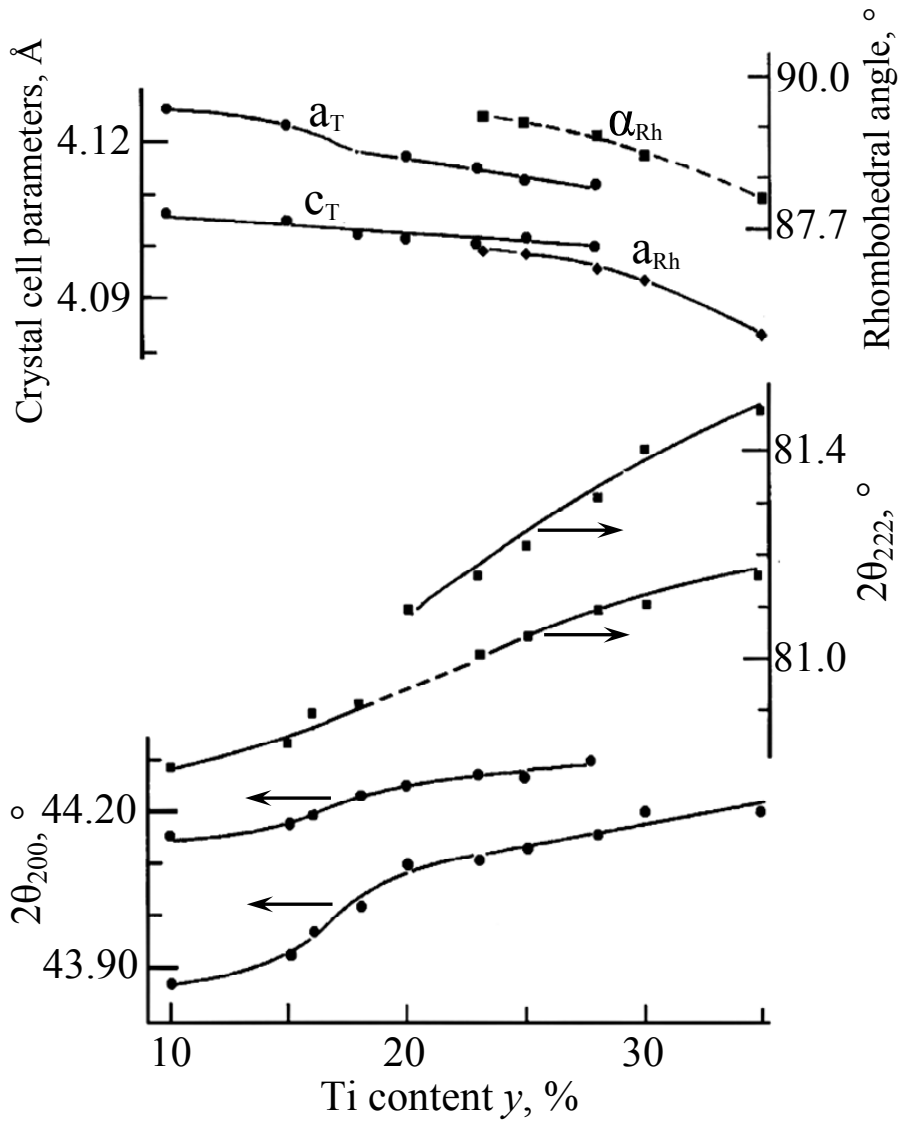

**Fig.6. Dependencies of the crystal cell parameters on Ti-content in 6/100−Y/Y PLZT series are shown at the top of the figure. The parameters $a_T$, and $c_T$ correspond to the tetragonal (AFE) phase and the crystal cell parameter $a_{Rh}$, and the $\alpha_{Rh}$ rhombohedral angle are for rhombohedral (FE) phase. Dependencies of the positions (the $2\theta_{200}$ and $2\theta_{222}$ angles) of the peaks of components for the (200) and the (222) X-ray lines on Ti-content in 6/100−Y/Y PLZT are presented at the bottom of the figure.**

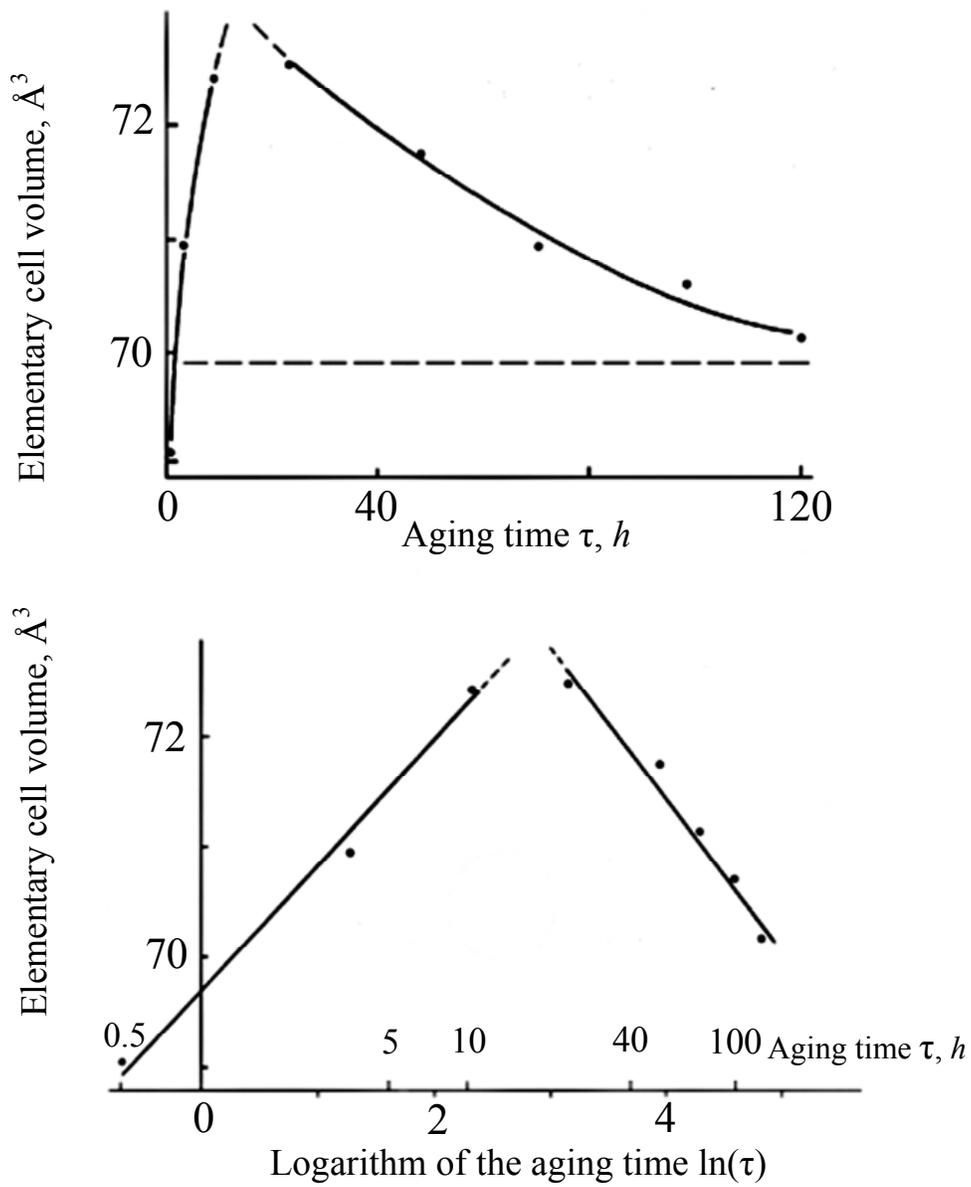

**Fig.7. Dependence of the elementary cell volume on the aging time for the 6/73/27 PLZT solid solution.**

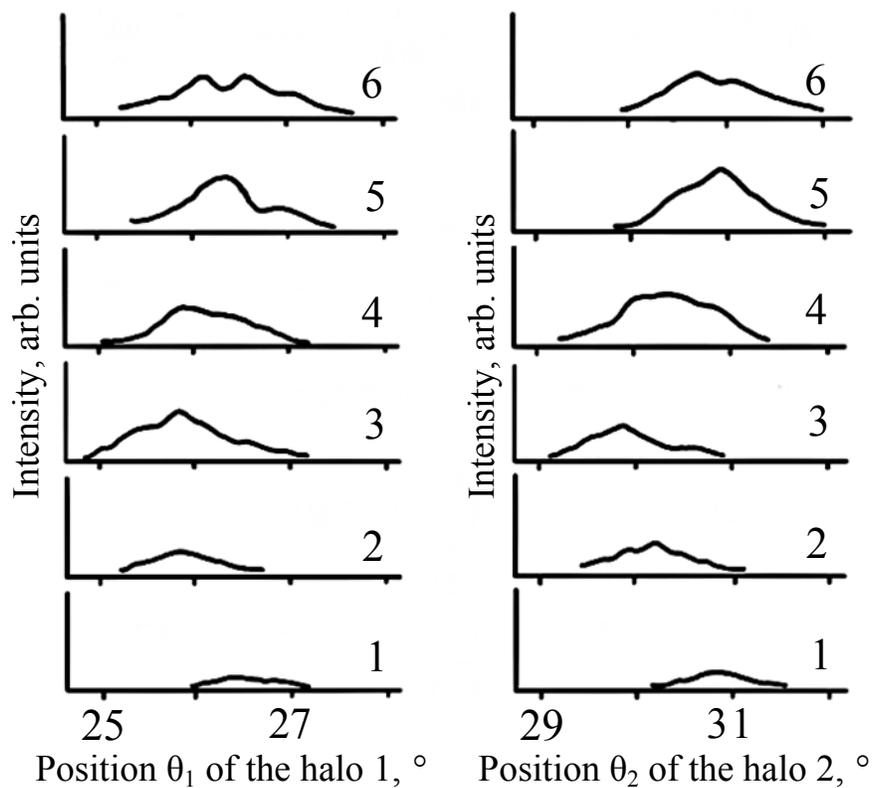

**Fig.8.** Changes of profile and position of the halo in the process of aging after the quenching of the 6/73/27 PLZT solid solution.

**Aging time (hours): 1 – 0.5, 2 – 3.5, 3 – 23, 4 – 48, 5 – 72, 6 – 120.**

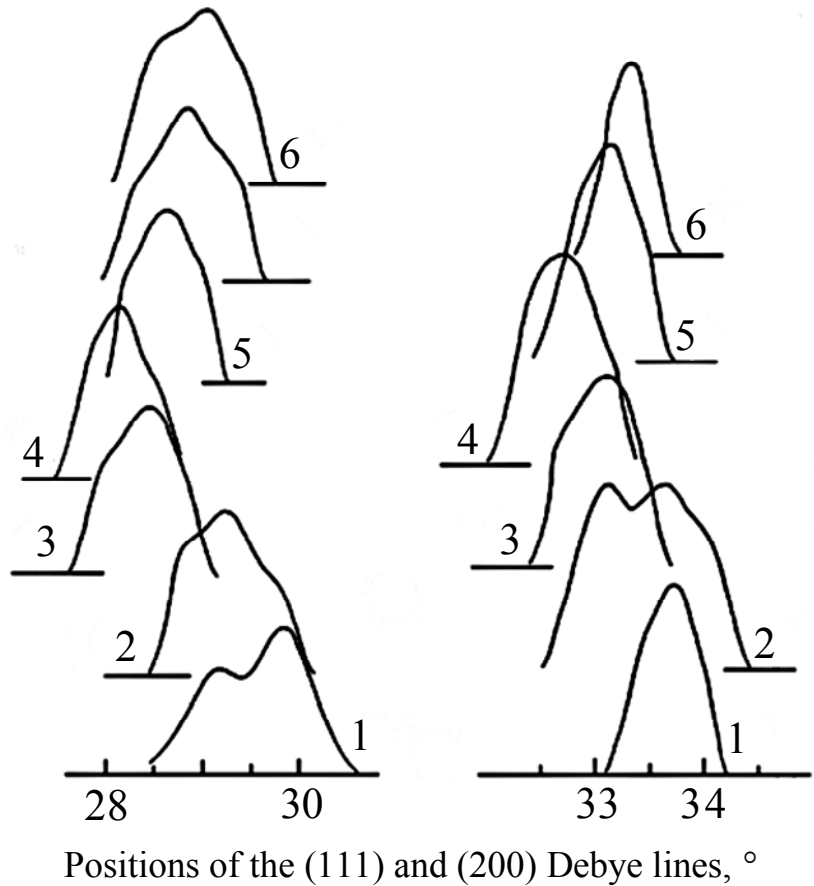

Positions of the (111) and (200) Debye lines, °

**Fig.9. Changes of profile and position of the (111) Debye line (left) and the (200) Debye line (right) in the process of aging after the quenching of the 6/73/27 PLZT solid solution.**

**Aging time (hours): 1 – 0.5, 2 – 3.5, 3 – 23, 4 – 48, 5 – 72, 6 – 96, 7 - 120.**

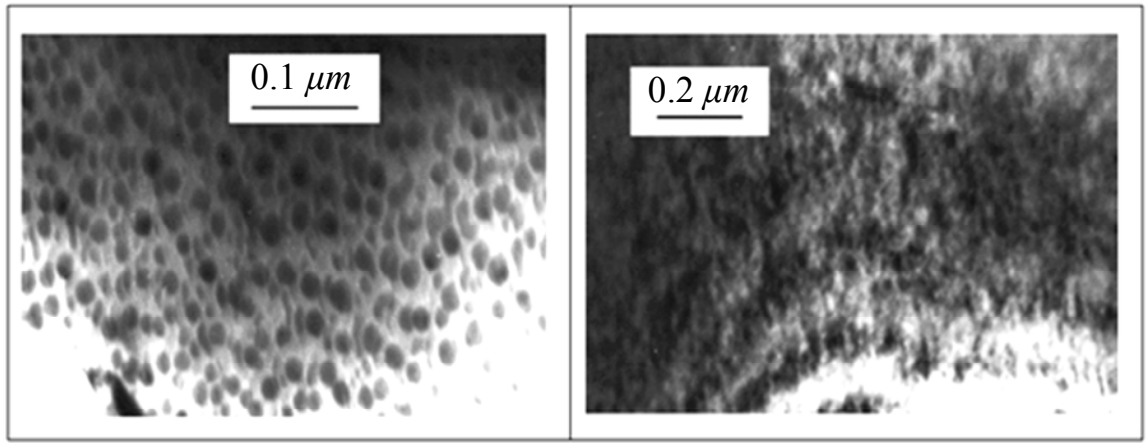

**Fig.10. TEM bright-field images for 7/65/35 PLZT solid solution. In the left – the electron beam is along the [110] direction. In the right – the electron beam is along the [120] direction.**